\documentclass[pop,fleqn]{w-art}
\usepackage{times}
\usepackage{amsfonts}
\usepackage{amsmath}
\usepackage{amssymb}
\usepackage[]{graphicx}
\chardef\bslash=`\\ 

\begin{document}
\DOIsuffix{theDOIsuffix}
\Volume{51}
\Issue{1}
\Month{01}
\Year{2003}
\pagespan{1}{}
\keywords{Black holes, closed timelike curves, G\"odel-type universes, holography.}
\subjclass[pacs]{04.60.-m, 04.70.-s, 04.70.Dy, 11.25.Tq}


\title[]{On the Thermodynamics of G\"odel Black Holes}

\author[]{Dietmar Klemm\inst{1,2}}
     \address[\inst{1}]{Universit\`{a} di Milano, Dipartimento di Fisica,
     Via Celoria 16, I-20133 Milano} \address[\inst{2}]{INFN, Sezione di Milano, Via Celoria 16, I-20133 Milano}
\author[]{Luciano Vanzo\inst{3,4}}
\address[\inst{3}]{Universit\`{a} di Trento, Dipartimento di Fisica, Via Sommarive 14, I-38050 Trento}
\address[\inst{4}]{INFN, Gruppo Collegato di Trento, Via Sommarive 14, I-38050 Trento}
\begin{abstract}
  After a brief review of G\"odel-type universes in string theory, we discuss
  some intriguing properties of black holes immersed in such backgrounds.
  Among these are the upper bound on the entropy that points towards a
  finite-dimensional Hilbert space of a holographically dual theory,
  and the minimum black hole temperature that is reminiscent of the Hawking-Page
  transition. Furthermore, we discuss several difficulties that are encountered
  when one tries to formulate a consistent thermodynamics of G\"odel black holes,
  and point out how they may be circumvented.
  \end{abstract}
\maketitle                   





\section{Introduction}

One of the most surprising results of the classification of all BPS solutions
to $N=2$ supergravity in five dimensions \cite{bib1} was the existence of a maximally
supersymmetric G\"odel-type universe, with metric and U$(1)$ gauge field
given by
\begin{equation}
  ds^2 = -(dt + j r^2\sigma_3)^2 + dr^2 + \frac{r^2}4(\sigma_1^2 + \sigma_2^2 + \sigma_3^2)\,, \qquad
  A = \frac{\sqrt 3}2 j r^2\sigma_3\,, \label{goedel}
\end{equation}
where
\begin{eqnarray}
\sigma_1 &=& \sin\phi d\theta - \cos\phi \sin\theta d\psi\,, \nonumber \\
\sigma_2 &=& \cos\phi d\theta + \sin\phi \sin\theta d\psi\,, \\
\sigma_3 &=& d\phi + \cos\theta d\psi\,. \nonumber
\end{eqnarray}
denote right-invariant one-forms on SU$(2)$, with Euler angles $(\theta,\phi,\psi)$.
$j$ is a real parameter
responsible for the rotation of (\ref{goedel}). This solution solves the equations
of motion following from the action
\begin{equation}
I = \frac 1{16\pi G}\int d^5 x \sqrt{-g} [R - F^2] - \frac 1{24\sqrt 3 \pi G}\int d^5 x\,
    \epsilon^{\mu\nu\alpha\beta\gamma} F_{\mu\nu} F_{\alpha\beta} A_{\gamma}\,. 
\end{equation}
Like the original G\"odel universe \cite{bib2}, presented by Kurt G\"odel in 1949 on
occasion of Albert Einstein's 70th birthday, this solution is homogeneous:
It has a nine-dimensional group of bosonic isometries
H$(2)$ $\ltimes$ (SU$(2)_R$ $\times$ U$(1)_L$) \cite{bib3}, where H$(2)$ denotes the
Heisenberg group with five generators. There are further common features of (\ref{goedel})
and its four-dimensional cousin: The stress tensor of the Maxwell field in (\ref{goedel})
is that of pressureless dust with energy density proportional to $j^2$ \cite{bib1}.
In addition, just like the original G\"odel universe, the solution (\ref{goedel}) suffers
from closed timelike curves (CTCs); the induced metric on hypersurfaces of constant
$t$ and $r$ becomes Lorentzian for $r > 1/j$.

One can lift the five-dimensional G\"odel-type universe to $D=11$ supergravity.
The solution is then just a product spacetime of (\ref{goedel}) and a six-dimensional
flat space $\mathbb{R}^6$. This configuration is highly supersymmetric, it preserves 20
supercharges \cite{bib1}. Let us denote the coordinates of $\mathbb{R}$ by $y^a$.
Dimensional reduction along say $y^6$ and subsequent T-duality along $y^5$ yields then
the type IIB pp-wave resulting from the Penrose limit of
AdS$_3$ $\times$ S$^3$ $\times$ T$^4$ \cite{bib3}.
Actually what one gets is a compactified pp-wave, whose CTCs can be eliminated by going to the
universal covering space. Note that this is not possible for the solution (\ref{goedel}),
which is already topologically trivial.
 
The appearance of CTCs creates of course several pathologies, for instance the
possibility of time-travel or the fact that the Cauchy problem in such spacetimes is
always ill-defined, to mention only two of them.
There is an ongoing research activity on G\"odel-type universes,
with the general aim to shed light on the status of closed timelike curves
in string theory. One would like to understand how string theory resolves
the pathologies mentioned above, and if we should discard solutions like (\ref{goedel}),
in spite of their high amount of supersymmetry. We will not discuss these issues here,
and refer the reader to \cite{bib3,bib4,bib5,bib6} and references therein.
Apart from questions related to causality problems, the interest in G\"odel-type
spacetimes is also motivated by holography: Similar to de~Sitter spaces, the
G\"odel universe shows observer-dependent holographic screens \cite{bib3},
and there is evidence for a finite-dimensional Hilbert space of a putative
holographically dual theory. The G\"odel-type universe represents thus a
supersymmetric laboratory for addressing conceptual puzzles of de~Sitter
holography \cite{bib3}. As we will see, of particular interest in these contexts
are black holes immersed in G\"odel universes, which we will discuss in the remainder
of this paper.

\section{Black Holes in G\"odel Universes}

Gimon and Hashimoto presented a solution that can be viewed as a Kerr black hole
(with two equal rotation parameters $l_1 = l_2 \equiv l$) embedded in the
five-dimensional G\"odel universe (\ref{goedel})\footnote{An extremal supersymmetric
Reissner-Nordstr\"om-G\"odel black hole was found in \cite{bib1}, and its
properties were studied in \cite{bib6}.}. The metric and gauge field are given by \cite{bib7}
\begin{equation}
ds^2 = -f(r)\left(dt + \frac{a(r)}{f(r)}\sigma_3\right)^2 + \frac{dr^2}{V(r)} +
       \frac{r^2}4(\sigma_1^2 + \sigma_2^2) + \frac{r^2V(r)}{4f(r)}\sigma_3^2\,,
       \qquad A = \frac{\sqrt 3}2 j r^2 \sigma_3\,, \label{kerrgoedel}
\end{equation}
where
\begin{equation}
f(r) = 1 - \frac{2m}{r^2}\,, \qquad a(r) = jr^2 + \frac{ml}{r^2}\,, \qquad
V(r) = 1 - \frac{2m}{r^2} + \frac{16j^2m^2}{r^2} + \frac{8jml}{r^2} + \frac{2ml^2}{r^4}\,.
\end{equation}
$m$ represents the "mass" parameter of the solution. For $j=0$, (\ref{kerrgoedel})
reduces to the Kerr black hole with equal rotation parameters. For $l=m=0$,
we recover the G\"odel universe (\ref{goedel}). In what follows, we shall consider
only the case $l=0$, i.~e.~, the Schwarz\-schild-G\"odel black hole.
This has an event horizon at $r_H = \sqrt{2m(1 - 8j^2m)}$ and an ergosphere at
$r_{erg} = \sqrt{2m}$. For $r = r_{CTC} = \sqrt{1 - 8j^2m}/2j$
(the velocity of light surface), $\partial_{\phi}$ becomes lightlike.
If $r > r_{CTC}$, $\partial_{\phi}$ is timelike and CTCs appear.
Note that $r_H < r_{CTC}$ for $0 < m < 1/8j^2$. $m_{max} = 1/8j^2$ is the
maximal mass parameter; for $m > m_{max}$ there is no horizon. Notice also
that for $m \neq 0$, the electromagnetic stress tensor can no more be written
in perfect fluid form.

The vector field
\begin{equation}
{\cal N}u = \partial_t + \frac{4j}{1 - 4j^2r^2 - 8j^2m}\partial_{\phi}
\end{equation}
is orthogonal to the natural time slicing determined by $t$ and becomes null on
the horizon. $u$ is the future directed normal field and the lapse
\begin{equation}
{\cal N}^2 = \frac{r^2 - 2m + 16j^2m^2}{r^2(1 - 4j^2r^2 - 8j^2m)}
\end{equation}
vanishes on the horizon but is infinite on $r_{CTC}$. The horizon limit field
\begin{equation}
K = \partial_t + \omega_H \partial_{\phi}\,, \qquad \omega_H = \frac{4j}{(1 - 8j^2m)^2}\,,
\end{equation}
is a timelike Killing field for $r \ge r_H$. Note that the horizon angular velocity
$\omega_H$ is nonzero because we have a black hole in a rotating universe rather
than a rotating black hole that lives in a static background.

\begin{figure}[htb]
\begin{minipage}[t]{.45\textwidth}
\includegraphics[width=\textwidth]{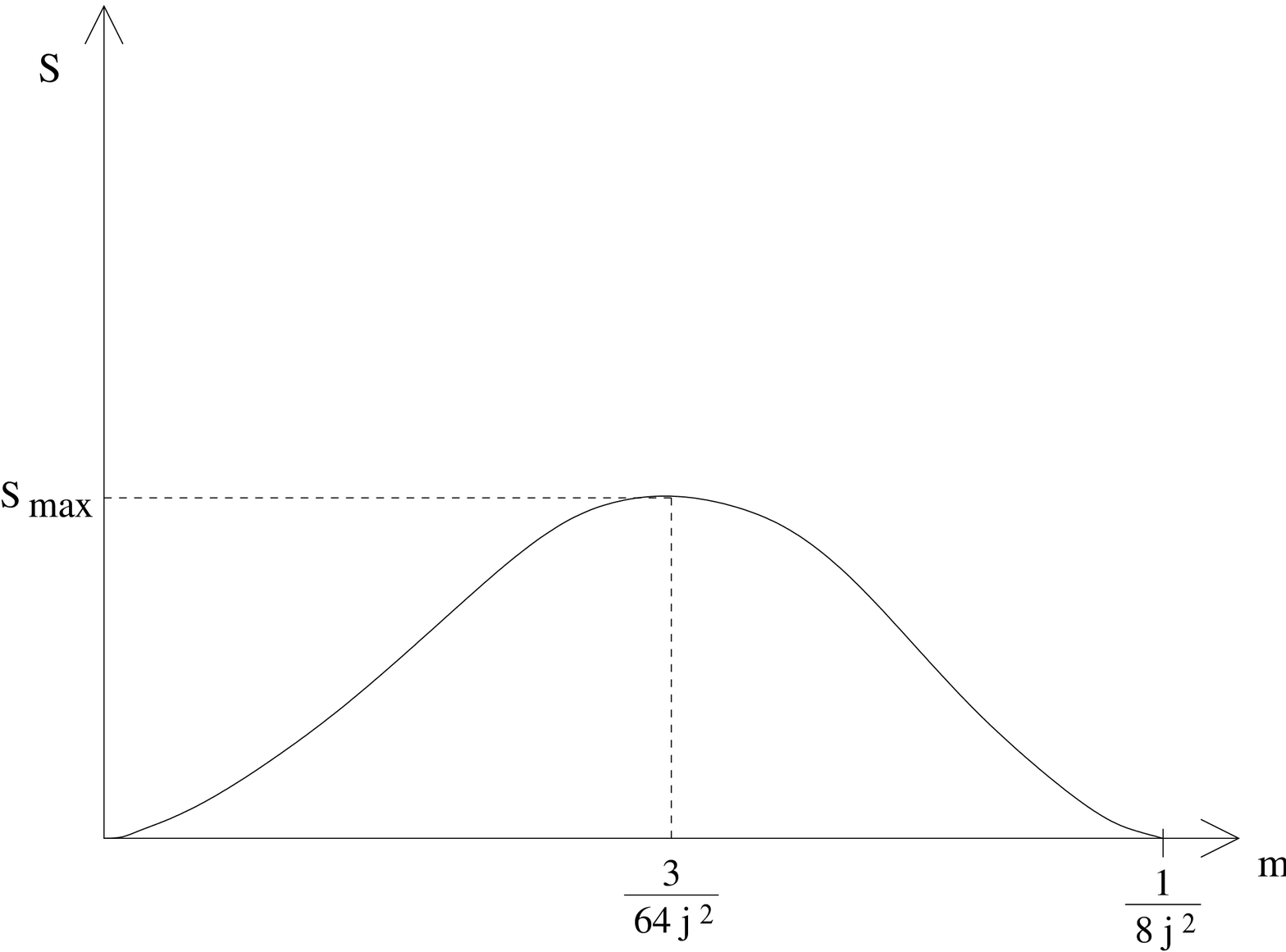}
\caption{Entropy $S$ of the Schwarz\-schild-G\"odel black hole
as a function of the mass parameter.}
\label{S}
\end{minipage}
\hfil
\begin{minipage}[t]{.45\textwidth}
\includegraphics[width=\textwidth]{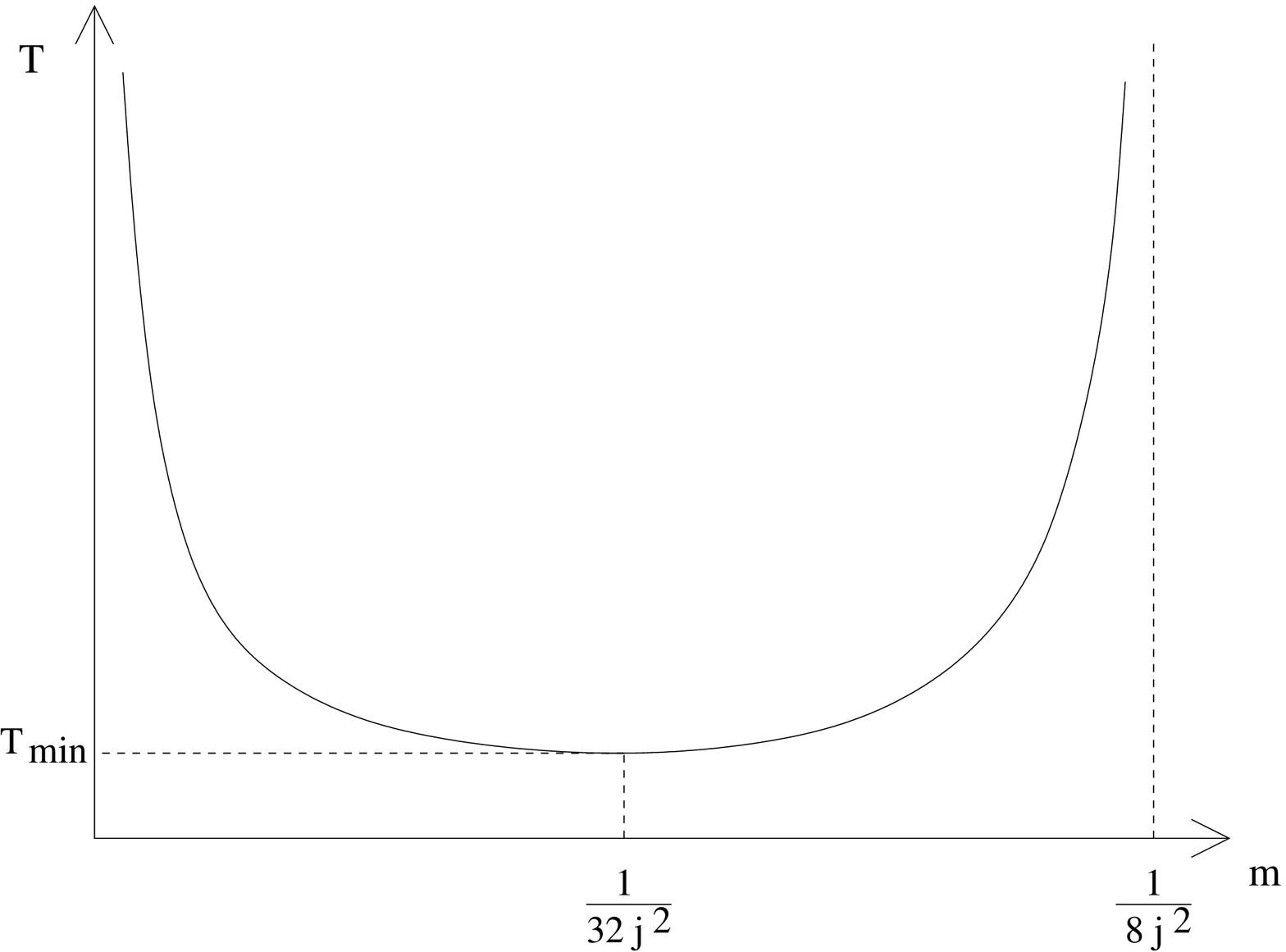}
\caption{Temperature $T$ as a function of $m$.}
\label{T}
\end{minipage}
\end{figure}

It is tempting to associate to this black hole an entropy
\begin{equation}
S = \frac{A_H}{4G} = \frac{\pi^2}{2G}\sqrt{8m^3(1 - 8j^2m)^5}\,.
    \label{entropy}
\end{equation}
Figure \ref{S} shows $S$ as a function of the mass parameter $m$.
We see that, in contrast to black holes in asymptotically flat spaces or in AdS,
G\"odel black holes cannot have an arbitrarily large entropy. $S$ becomes
maximal for $m = 3/64j^2$ and decreases when $m$ is further increased.
As the Schwarz\-schild-G\"odel black hole represents a finite temperature
excitation above the supersymmetric G\"odel ground state, it seems that there
is an upper bound on the entropy of such excitations. Similar to the case of
de~Sitter gravity, whose holographic dual was argued to have a finite number
of states \cite{bib8} (cf.~also \cite{bib9} for a recent review),
this seems to point towards a finite-dimensional Hilbert
space of the quantum description. In de~Sitter space, the entropy of a black
hole is bounded from above by the entropy of the Nariai solution, which is
the largest black hole that can fit within the cosmological horizon.
Something similar happens for the G\"odel black hole, where the surface
$r = r_{CTC}$ (beyond which CTCs appear) acts effectively like a box:
The three-sphere that is tangent to the horizon develops timelike directions
for $r > r_{CTC}$, but timelike vectors cannot be tangent to null surfaces,
so there can be no horizons for $r > r_{CTC}$.
That's why there is an upper bound on the mass parameter and on the entropy
of the G\"odel black hole.

The horizon temperature is given in terms of the surface gravity $\kappa$,
\begin{equation}
T = \frac{\kappa}{2\pi} = \frac 1{2\pi\sqrt{2m(1 - 8j^2m)^3}}\,.
\end{equation}
Alternatively, it can be found by demanding smoothness
of the Euclidean section, which is obtained by the analytic
continuation
\begin{equation}
t \to \tau = -it\,, \qquad j \to J = ij\,.
\end{equation}
This implies that the Euclidean time $\tau$ is identified modulo $\beta = 1/T$,
and $\phi \sim \phi + \beta\Omega_H$, where
\begin{equation}
\Omega_H = \frac{4J}{(1 + 8J^2m)^2}\,.
\end{equation}
Quite surprisingly, it seems that the Euclidean section is perfectly
well-behaved in the whole range of $r$, in spite of the pathologies
encountered in its Lorentzian counterpart.

Figure \ref{T} shows $T$ as a function of $m$. Obviously there
is a minimum temperature $T_{min} = 16j/(3\sqrt 3\pi)$ for $m = 1/32j^2$,
so that the Schwarz\-schild-G\"odel black hole can exist only above this
temperature. This is reminiscent of Schwarz\-schild-AdS black holes,
where a similar behaviour leads to the Hawking-Page phase transition.

In order to study more in detail the thermodynamics of the G\"odel black
hole, one needs a notion of mass and angular momentum in G\"odel
universes. It is clear that $m$ and $j$ are not the thermodynamical
parameters, because the partial derivative of the entropy (\ref{entropy})
with respect to $m$, keeping $j$ fixed, does not yield the correct inverse temperature.
One possible way to define a mass is by means of the Euclidean
action. It is straightforward to show that in our case, the on-shell
Euclidean action reduces to the boundary term
\begin{equation}
I_E = \frac 1{12\pi G}\int_{\partial{\cal M}} d^4 S_{\mu} A_{\nu} F^{\mu\nu}
      - \frac 1{8\pi G}\int_{\partial{\cal M}} d^4 x\sqrt h K\,,
\end{equation}
where $d^4 S_{\mu} = n_{\mu} d^4 x$ with $n_{\mu}$ the outward pointing
unit normal to the boundary $\partial{\cal M}$, $h_{\mu\nu}$ is the induced metric
on the boundary, and $K$ denotes the trace of its extrinsic curvature. The last term
is the Gibbons-Hawking boundary term necessary in order to have a well-defined
variational principle. Taking $\partial{\cal M}$ to be a surface of
constant $r$, one finds
\begin{equation}
I_E = \frac{\pi\beta}{32 G}(3r^2 + 8J^2 m r^2 - 4m - 32J^2 m^2 - 4J^2 r^4)\,.
\end{equation}
We were not able to regularize the divergence in $I_E$ for $r \to \infty$ by
subtracting a suitable background. The obvious background would be the solution
with $m = 0$, i.~e.~, the supersymmetric G\"odel universe. However, strictly
speaking, the G\"odel black hole does not approach the geometry with $m=0$
asymptotically, because
\begin{equation}
g_{\mu\nu} - g_{\mu\nu}^{m = 0} \sim - 2j^2 m r^2 \sigma_3^2\qquad {\mathrm{for}}
\quad r \to \infty\,.
\end{equation}
The absence of a proper background to which the spacetime asymptotes is also one
of the reasons that the usual Abbott-Deser construction to compute the mass and
angular momentum does not work here. (Another reason is the fact that the conserved
charges calculated with the Abbott-Deser construction involve surface
integrals at infinity, but, as the induced metric on such surfaces becomes
Lorentzian due to the presence of CTCs, these integrals would be imaginary).
A possible way out could be to put the G\"odel black hole in a box $R < r_{CTC}$,
and to compute the stress-energy-momentum content of the bounded spacetime
region $r \le R$ using the Brown-York formalism \cite{bib10}.
In the case of the four-dimensional Schwarz\-schild black hole, this yields
a first law in the form \cite{bib10}
\begin{equation}
dE = p dV + T_{local} dS\,,
\end{equation}
where
\begin{equation}
T_{local} = (8\pi M\sqrt{1 - 2M/R})^{-1}
\end{equation}
is the local Tolman temperature, blueshiftet from infinity to a finite radius $R$.
Note the appearance of the additional term $pdV$ in the first law, which comes
from the fact that one is considering a finite volume $V = 4\pi R^2$.
One can now try to generalize this to the case of the Schwarz\-schild-G\"odel
black hole. Work in this direction is in progress.

Alternatively, one can proceed as in \cite{bib11} to obtain a mass formula by
integrating the Killing identity
\begin{equation}
\nabla_{\nu}\nabla_{\mu}\xi^{\nu} = R_{\mu\sigma}\xi^{\sigma} =
2(F_{\mu\nu}{F_{\sigma}}^{\nu} - \frac 16 g_{\mu\sigma} F^2)\xi^{\sigma}\,,
\end{equation}
where $\xi$ is a Killing vector, on a spacelike hypersurface $\Sigma_t$
of constant $t$ from the black hole horizon $r_H$ to some $R < r_{CTC}$.
Using $\xi = \partial_t$ we obtain for the total mass $M$ contained
in the region $r \le R$ the Smarr-like formula
\begin{equation}
M = \frac 3{16\pi G}\kappa A_H + \frac 32 \omega_H J_H + \frac 32
    \int_{\Sigma_t}(T_{\mu\sigma} - \frac 13 T g_{\mu\sigma})\xi^{\sigma}d\Sigma^{\mu}\,,
    \label{smarr}
\end{equation}
where
\begin{equation}
J_H = -\frac 1{16\pi G}\int_{\mathrm{Hor}}\nabla_{\mu}{\tilde K}_{\nu}d\Sigma^{\mu\nu}\,,
      \qquad \tilde K = \partial_{\phi}\,,
\end{equation}
is the angular momentum of the horizon, and
\begin{equation}
T_{\mu\sigma} = \frac 1{4\pi G}(F_{\mu\nu}{F_{\sigma}}^{\nu} - \frac 14 g_{\mu\sigma}F^2)
\end{equation}
denotes the stress tensor of the electromagnetic field. The last term in (\ref{smarr})
represents the contribution of the gauge field to the energy in the region $r \le R$.
Evaluating (\ref{smarr}) yields
\begin{equation}
M = \frac{3\pi}{4G}[m(1 - 8j^2m) + 2j^2 R^2(R^2 - 2m)]\,. \label{smarreval}
\end{equation}

Further difficulties are related to the different notions of charge that
appear in presence of Chern-Simons terms \cite{bib12}. To see this, consider the
Maxwell equations
\begin{equation}
\nabla_{\mu} F^{\mu\nu} = 4\pi J^{\nu}_{ext}\,,
\end{equation}
where $J_{ext}$ denotes an external source. Integrating this on a spacelike
hypersurface $\Sigma$ and using Gauss' law yields
\begin{equation}
Q_M = \frac 1{4\pi}\int_{\partial\Sigma} F^{\mu\nu} d\Sigma_{\mu\nu} \label{maxwcharge}
\end{equation}
for the "Maxwell charge" $Q_M = \int_{\Sigma} J^{\nu}_{ext} d\Sigma_{\nu}$. (\ref{maxwcharge})
diverges for the G\"odel black hole, if $\partial\Sigma$ is a surface of constant $t,r$
and $r \to \infty$.

In our case, we have a Chern-Simons term, and thus the Maxwell equations
are modified to
\begin{equation}
\nabla_{\mu} (F^{\mu\nu} + J^{\mu\nu}) = 4\pi J^{\nu}_{ext}\,,
\end{equation}
with the Chern-Simons current
\begin{equation}
J^{\mu\nu} = \frac 1{\sqrt 3}\epsilon^{\mu\nu\alpha\beta\gamma}A_{\alpha}F_{\beta\gamma}\,.
\end{equation}
Proceeding as before yields the "Page charge"
\begin{equation}
Q_P = \frac 1{4\pi}\int_{\partial\Sigma}(F^{\mu\nu} + J^{\mu\nu}) d\Sigma_{\mu\nu}\,,
\end{equation}
which vanishes for the G\"odel black hole. These different notions of charge
raise the question which one enters the first law of black hole mechanics.

If we use the Page charge, which takes into account only external sources,
but not the contributions that come from the Chern-Simons current $J^{\mu\nu}$,
the Schwarz\-schild-G\"odel solution is not electrically charged.
It might possess, however, a magnetic dipole moment $\mu$.
Something similar happens for supertubes, to which the G\"odel spacetimes
are closely related \cite{bib5}. For instance the type IIA supertube carries no
D2-brane charge, but it does have a D2-dipole moment \cite{bib13}.
The question is thus if there appears a term $B d\mu$ in the first law
of the G\"odel black hole ($B$ denotes the magnetic field). As far as we are
aware, up to now there are no known examples of black holes that have such a
term in the first law\footnote{The black rings of \cite{bib14} carry a dipole moment
that is independent of their electric charge, but the thermodynamics of these
objects has not yet been worked out.}. To determine $\mu$
is nontrivial; the magnetic dipole moment cannot be simply read off from the
asymptotic behaviour of the electromagnetic field, because the solution is
not asymptotically flat.

Adopting the notion of Page charge and assuming the absence of a magnetic
dipole moment, the first law for the G\"odel black hole should have
the form
\begin{equation}
dE = TdS + \omega_H dJ\,. \label{1stlaw}
\end{equation}
This can be integrated to get expressions for $E$ and $J$. For dimensional reasons
it is clear that $E$ and $J$ must have the form
\begin{equation}
E = \frac mG f(x)\,, \qquad J = \frac{jm^2}G h(x)\,,
\end{equation}
where $f$, $h$ are functions of $x \equiv j^2 m$ only. The first law (\ref{1stlaw})
implies then
\begin{eqnarray}
E &=& \frac{3\pi m}{4Gx}(1 - 8x)\left[\frac 1{24} + 4C(1 - 8x)^3\right]\,, \\
J &=& \frac{\pi j m^2}{32Gx^2}(1 - 8x)^3\left[\frac 16 - 4x + 16C(1 - 8x)^3\right]\,,
      \nonumber
\end{eqnarray}
with $C$ denoting an integration constant. Notice that it is not possible to
choose $C$ such that $E$ coincides with the total mass within the velocity
of light surface $r = r_{CTC}$, obtained by choosing $R = r_{CTC}$ in the Smarr
formula (\ref{smarr}), (\ref{smarreval}). Furthermore, for no choice of $C$
does $E$ reduce to the total mass within the holographic Bousso screen
(located at $r = \sqrt{3(1 - 8j^2m)}/4j$ \cite{bib7}), or to the expression
$3\kappa A_H/(16\pi G) + 3\omega_H J_H/2$ in (\ref{smarr}),
that represents somehow the mass of the black hole (being the total mass
minus the contribution of the gauge field). It is easy to see that also
$J \neq J_H$ for all values of $C$. This means that the mass $E$ and the
angular momentum $J$ that enter the first law must also have contributions from
the electromagnetic field.
Requiring $E$ to reduce to its correct value $E = 3\pi m/4G$ in the limit
$j \to 0$ fixes $C = -1/96$.

\section{Final Remarks}

We saw that black holes in G\"odel universes have some interesting
properties, like the upper bound on the entropy, that points towards a
finite-dimensional Hilbert space of the quantum description, or the minimum
temperature that is reminiscent of Schwarz\-schild-AdS black holes.
For the latter, the minimum temperature leads to the Hawking-Page
phase transition, but in the G\"odel case it is difficult to imagine
something similar, because a phase transition cannot occur in systems
with a finite number of degrees of freedom.

Perhaps a detailed analysis of the thermodynamical behaviour of
G\"odel black holes can shed more light on these questions.
As we saw, the formulation of a consistent thermodynamics is
hindered by the difficulties that one encounters in defining
conserved charges like mass and angular momentum. These problems
are intimately related to the appearance of CTCs.

One way to overcome them may be to embed the
G\"odel black hole in an AdS background\footnote{Unfortunately the
corresponding metric is not yet known.} and to study the
thermodynamics of this "Schwarz\-schild-G\"odel-AdS black hole".
The motivation for this rests on the fact that a negative cosmological
constant $\Lambda$ can act as a regulator for CTCs: In the G\"odel-AdS
solution found in \cite{bib15}, all CTCs disappear if the absolute value
of $\Lambda$ is large enough compared to the G\"odel parameter
$j$ \cite{bib16}. The difficulties in defining conserved charges and
formulating a consistent thermodynamics, present
in the case $\Lambda = 0$ and discussed in detail above, might then
disappear. Having determined the thermodynamic potentials, one
can then study the limit when the cosmological constant approaches its
critical value where CTCs show up, and see whether at that point
phase transitions appear in the system.
We leave an analysis of these issues for a future publication.

\begin{acknowledgement}
  This work was partially supported by INFN, MURST and by the European Commission
  program MRTN-CT-2004-005104. We are grateful to S.~Cacciatori and C.~A.~R.~Herdeiro
  for valuable discussions.
\end{acknowledgement}

\end{document}